\documentclass[letterpaper,superscriptaddress,preprint,byrevtex,showpacs,prb]{revtex4-1}

\DeclareSymbolFont{lettersA}{U}{pxmia}{m}{it}
\DeclareMathAlphabet{\mathsfsl}{OT1}{cmss}{m}{sl}

\SetSymbolFont{lettersA}{bold}{U}{pxmia}{bx}{it}
\DeclareFontSubstitution{U}{pxmia}{m}{it}
\DeclareSymbolFontAlphabet{\mathfrak}{lettersA}
\DeclareMathSymbol{\piup}{\mathord}{lettersA}{"19}
\DeclareMathSymbol{\iTheta}{\mathalpha}{letters}{2}

\usepackage{amsmath}
\usepackage{amssymb}
\usepackage{multirow}
\usepackage{graphicx}
\usepackage{epstopdf}
\usepackage[hang,scriptsize]{subfigure}
\usepackage{bbm}
\usepackage{mathrsfs}
\usepackage{xcolor}
\usepackage[colorlinks=true,pdfstartview=FitV,linkcolor=blue,citecolor=blue,urlcolor=blue]{hyperref}

\makeatletter

\newcommand{\Rmnum}[1]{\expandafter\@slowromancap\romannumeral #1@}
\makeatother

\newcommand{\ii}{\mathrm{i}}
\newcommand{\diff}{\mathrm{d}}

\newcommand{\blue}[1]{\textcolor{blue}{#1}}

\begin{document}

\title{Nonlocal effective medium analysis in symmetric metal-dielectric multilayer metamaterials}

\author{Lei~Sun}
\address{Department of Mechanical and Aerospace Engineering,
    Missouri University of Science and Technology,
    Rolla, Missouri 65409, USA}

\author{Zhigang~Li}
\address{Department of Mechanical and Aerospace Engineering,
    Missouri University of Science and Technology,
    Rolla, Missouri 65409, USA}

\author{Ting~S.~Luk}
\address{Center for Integrated Nanotechnologies,
    Sandia National Laboratories,
    Albuquerque, New Mexico 87185, USA}

\author{Xiaodong~Yang}
\email[To whom all correspondence should be addressed: ]{yangxia@mst.edu}
\address{Department of Mechanical and Aerospace Engineering,
    Missouri University of Science and Technology,
    Rolla, Missouri 65409, USA}

\author{Jie~Gao}
\email[To whom all correspondence should be addressed: ]{gaojie@mst.edu}
\address{Department of Mechanical and Aerospace Engineering,
    Missouri University of Science and Technology,
    Rolla, Missouri 65409, USA}

\begin{abstract}
The optical nonlocality in symmetric metal-dielectric multilayer metamaterials is
theoretically and experimentally investigated with respect to the TM-polarized incident light.
A new nonlocal effective medium theory is derived from the transfer-matrix method to determine
the nonlocal effective permittivity depending on both frequency and wave vector in the symmetric
metal-dielectric multilayer stack.
In contrast to the local effective medium theory, our proposed nonlocal effective medium theory
can accurately predict the measured incident angle-dependent reflection spectra from the fabricated
multilayer stack and provide nonlocal dispersion relations.
Moreover, the bulk plasmon polaritons with large wave vectors supported in the multilayer
stack are also investigated with the nonlocal effective medium theory through the analysis
of dispersion relation and eigenmode.
\end{abstract}

\pacs{42.25.Bs, 78.20.Ci, 78.67.Pt, 81.05.Zx}

\maketitle

\section{Introduction}

Metal-dielectric multilayer metamaterials have recently emerged into the focus of extensive
exploration due to their anomalous electromagnetic properties in optical frequency range and
the straightforward fabrication process.
Metal-dielectric multilayer metamaterials with hyperbolic (or indefinite) dispersion have been
demonstrated to realize a broad range of applications \cite{Poddubny2013NP},
such as enhanced electromagnetic density of states \cite{Noginov2010OL,Jacob2010APB},
negative refraction \cite{Chui2006JP,Hoffman2007NM,Zhao2014AIP},
deep-subwavelength imaging \cite{Jacob2006OE,Liu2007Sci,Zhang2008NM},
spontaneous emission enhancement \cite{Jacob2012APL,Iorsh2012PLA,Lu2014NN},
thermal emission engineering \cite{Guo2012APL},
and anomalous indefinite cavities \cite{Yang2012NP}.
Furthermore, the metal-dielectric multilayer stack is also utilized to construct the
epsilon-near-zero (ENZ) metamaterials, which is of great interests in many research areas,
including radiation wavefront tailoring \cite{Alu2007PRB,Sun2013PRB,Sun2013APL},
invisible cloaking \cite{Alu2005PRE,Pendry2006Sci},
displacement current insulation \cite{Engheta2007Sci,Silveirinha2009PRL},
optical nonlinearity enhancement \cite{Argyropoulos2012PRB},
harmonic generation \cite{Vincenti2011PRA,Ciattoni2012PRA},
enhanced photonic density of states \cite{Vesseur2013PRL},
and soliton excitations \cite{Rizza2011PRA}.
The electromagnetic properties of the metal-dielectric multilayer stack are simply
characterized by the local effective medium theory (EMT) since the nanoscale multilayer
period is much smaller than the electromagnetic wavelength.
In fact, the variation of the electromagnetic field on the scale of the multilayer
period will result in the spatial dispersion, leading to the optical nonlocality \cite{Elser2007APL},
which has been studied in other types of metamaterials such as split-ring resonator arrays \cite{Belov2005PRE}
and nanorod structures \cite{Silveirinha2006PRE,Belov2003PRB}.
Due to the strong optical nonlocality in the metal-dielectric multilayer stack,
especially when the frequency of the electromagnetic field approaches to the ENZ position \cite{Sun2013OE},
several extraordinary optical phenomena appear, such as the additional light waves \cite{Orlov2011PRB}
and complex eigenmodes \cite{Orlov2013OE}, which cannot be predicted by the local EMT.
In order to address this issue, the limitation of the local EMT has been studied recently \cite{Kidwai2012PRA},
and several different nonlocal EMT models have been proposed,
such as the field averaging algorithm confined to the lossless condition \cite{Chebykin2011PRB,Chebykin2012PRB}
and the dispersion relation approximation limited to the normal incident light \cite{Gao2013APL}.

In this work, the optical nonlocality is theoretically and experimentally studied in the
symmetric metal-dielectric multilayer stack with respect to the TM-polarized incident light
for different incident angles.
The optical nonlocality for the TM-polarized incident light is much stronger than that for
the TE-polarized light that we have studied previously \cite{Sun2014OE},
due to the fact that the effective permittivity tensor for the TM-polarized light shows strong
anisotropy.
Here a new nonlocal EMT is derived based on the original definition of the effective permittivity
through the transfer-matrix method \cite{Born1999} in order to analytically describe
the variation of the electromagnetic field across the symmetric metal-dielectric multilayer stack
with respect to both frequency and wave vector.
It is demonstrated that the measured incident angle-dependent reflection spectra from the fabricated
multilayer stack can be predicted accurately by the proposed nonlocal EMT, instead of the local EMT.
Furthermore, the difference between the nonlocal effective permittivity and the local effective
permittivity is also analyzed in detail, together with the ENZ position shift and the variation
of iso-frequency contour (IFC) induced by the optical nonlocality.
Moreover, the bulk plasmon polaritons (BPPs)\cite{Avrutsky2007PRB,Sreekanth2013SR},
a sort of highly confined optical modes with large
wave vectors generated from the coupling of the surface plasmon polaritons (SPPs) propagating
along the interfaces of the metal-dielectric multilayer stack,
are also investigated with the nonlocal EMT through the analysis of dispersion relation and eigenmode.

\section{Development of Nonlocal Effective Medium Theory}

Figure~\blue{1(a)} illustrates the schematic of the symmetric metal-dielectric multilayer stack composed
of $4$-pair periodic silver ($\mathrm{Ag}$) and silica ($\mathrm{SiO}_{2}$) layers on the top of a thick
silver substrate.
The $\mathrm{Ag}$-$\mathrm{SiO}_{2}$ multilayer stack possesses a symmetric unit structure with one
half-thickness $\mathrm{SiO}_{2}$ layer on the top of the $\mathrm{Ag}$ layer and the other at the bottom.
The silver substrate acts as a mirror to block the transmission and enhance the reflection from the multilayer
stack so that the optical nonlocality is only strongly related to the measured reflection spectra in experiments.
The permittivity and the thickness of the $\mathrm{Ag}$ layer and the $\mathrm{SiO}_{2}$ layer are individually
denoted as ($\varepsilon_{m}$, $a_{m}$) and ($\varepsilon_{d}$, $a_{d}$), while the thickness of the silver substrate
is denoted as $a_{\mathrm{sub}}$.
Here the TM-polarized incident light propagating in the $x$-$z$ plane with an arbitrary incident angle $\theta_{0}$
is considered.
In general, the symmetric $\mathrm{Ag}$-$\mathrm{SiO}_{2}$ multilayer stack can be regarded as a bulk homogenous
and anisotropic effective medium on the top of the silver substrate as shown in Fig.~\blue{1(b)}.
With respect to the TM-polarized light propagating in the $x$-$z$ plane, the local anisotropic effective permittivity
of the multilayer stack can be approximated by the local EMT as
$\varepsilon_{x}^{\mathrm{loc}} = (\varepsilon_{m}a_{m} + \varepsilon_{d}a_{d})/(a_{m} + a_{d})$
and
$\varepsilon_{z}^{\mathrm{loc}} = \varepsilon_{m}\varepsilon_{d}(a_{m} + a_{d})/(\varepsilon_{m}a_{d} + \varepsilon_{d}a_{m})$.
Clearly, the local effective permittivity is only a function of frequency, without considering the spatial dispersion
caused by the optical nonlocality.
However, previous studies show that the metal-dielectric multilayer stack possesses strong optical nonlocality,
leading to the nonlocal effective permittivity that is not only related to the frequency but also to the wave vector.

In order to take into account the optical nonlocality, the nonlocal EMT based on the original definition of the
effective permittivity is proposed and derived through the transfer-matrix method, where the $\mathrm{Ag}$-$\mathrm{SiO}_{2}$
multilayer stack is considered as a one-dimensional photonic crystal structure.
In the uniaxial multilayer stack, the nonlocal effective permittivity tensor of the stack can be presented
as a diagonal matrix with non-zero diagonal components
$\varepsilon_{x}^{\mathrm{nonloc}}$ and $\varepsilon_{z}^{\mathrm{nonloc}}$,
while the off-diagonal components are negligible.
Furthermore, due to the symmetric and periodic property of the $\mathrm{Ag}$-$\mathrm{SiO}_{2}$ multilayer stack,
the nonlocal effective permittivity is independent of the number of layers.
Therefore, only one symmetric unit cell embedded in a homogenous and isotropic surrounding medium
is considered in the calculation of the nonlocal effective permittivity, as displayed in Fig.~\blue{1(c)}.
For the TM-polarized incident light, the electric field in each layer can be presented as a linear combination of
the forward propagating wave (along the positive $z$-direction) and the backward propagating wave (along the negative $z$-direction)
\begin{equation}
\label{eq:e-field}
    E_{i} = E_{i}^{+}\exp\left(\ii k_{z_{i}}z\right) + E_{i}^{-}\exp\left(-\ii k_{z_{i}}z\right),
\end{equation}
where $i$ represents the layer number shown in Fig.~\blue{1(c)} and $i=0,1,2,3,4$.
Note that the factor $\exp(\ii k_{x}x)$ in Eq.~\eqref{eq:e-field} is omitted since the wave vector $k_{x}$ along
the interfaces is preserved across each layer.
According to the boundary conditions, the electric fields across each interface are related via the transfer-matrix as
\begin{equation}
\label{eq:trans-maxtrix}
\begin{bmatrix}
    E_{i}^{+}\exp\left(\ii k_{z_{i}}d_{i}\right) \\
    E_{i}^{-}\exp\left(-\ii k_{z_{i}}d_{i}\right)
\end{bmatrix}
=
\begin{bmatrix}
    1/t_{i,i+1} &r_{i,i+1}/t_{i,i+1} \\
    r_{i,i+1}/t_{i,i+1} &1/t_{i,i+1}
\end{bmatrix}
\cdot
\begin{bmatrix}
    E_{i+1}^{+}\exp\left(\ii k_{z_{i+1}}d_{i}\right) \\
    E_{i+1}^{-}\exp\left(-\ii k_{z_{i+1}}d_{i}\right)
\end{bmatrix},
\end{equation}
in which the transmission coefficient $t_{i,i+1}$ and the reflection coefficient $r_{i,i+1}$ are
\begin{equation}
\label{eq:trans}
    t_{i,i+1} = \frac{2\sqrt{\varepsilon_{i}}\cos\theta_{i}}
        {\sqrt{\varepsilon_{i+1}}\cos\theta_{i} + \sqrt{\varepsilon_{i}}\cos\theta_{i+1}}
\end{equation}
and
\begin{equation}
\label{eq:refl}
    r_{i,i+1} = \frac{\sqrt{\varepsilon_{i+1}}\cos\theta_{i} - \sqrt{\varepsilon_{i}}\cos\theta_{i+1}}
        {\sqrt{\varepsilon_{i+1}}\cos\theta_{i} + \sqrt{\varepsilon_{i}}\cos\theta_{i+1}}
\end{equation}
with respect to the TM-polarized incident light.
Furthermore, the phase variation of the electric field in each layer can be determined as
\begin{equation}
\label{eq:int-phase}
    \int_{d_{i-1}}^{d_{i}}
    \left[ E_{i}^{+}\exp\left(\ii k_{z_{i}}z\right)
    + E_{i}^{-}\exp\left(-\ii k_{z_{i}}z\right) \right]\diff z
    = E_{i}^{+}\Delta\phi_{i}^{+} + E_{i}^{-}\Delta\phi_{i}^{-},
\end{equation}
where the phase factor $\Delta\phi_{i}^{+}$ and $\Delta\phi_{i}^{-}$ are expressed as
\begin{equation}
\label{eq:phase-fr}
    \Delta\phi_{i}^{+} = \frac{\exp(\ii k_{z_{i}}d_{i}) - \exp(\ii k_{z_{i}}d_{i-1})}
        {\ii k_{z_{i}}}
\end{equation}
and
\begin{equation}
\label{eq:phase-bk}
    \Delta\phi_{i}^{-} = \frac{\ii\left[\exp(-\ii k_{z_{i}}d_{i}) - \exp(-\ii k_{z_{i}}d_{i-1})\right]}
        {k_{z_{i}}}.
\end{equation}
Therefore, the nonlocal effective permittivity can be obtained based on the original definition of effective permittivity as
\begin{equation}
\label{eq:eps_x}
    \varepsilon_{x}^{\mathrm{nonloc}} = \frac{\langle D_{i} \rangle_{x}}{\langle E_{i} \rangle_{x}}
        = \frac{ \sum_{i=1}^{3} \varepsilon_{i}(E_{i}^{+}\Delta\phi_{i}^{+}+E_{i}^{-}\Delta\phi_{i}^{-})\cos\theta_{i} }
        { \sum_{i=1}^{3} (E_{i}^{+}\Delta\phi_{i}^{+}+E_{i}^{-}\Delta\phi_{i}^{-})\cos\theta_{i} }
\end{equation}
and
\begin{equation}
\label{eq:eps_z}
    \varepsilon_{z}^{\mathrm{nonloc}} = \frac{\langle D_{i} \rangle_{z}}{\langle E_{i} \rangle_{z}}
        = \frac{ \sum_{i=1}^{3} \varepsilon_{i}(E_{i}^{+}\Delta\phi_{i}^{+}+E_{i}^{-}\Delta\phi_{i}^{-})\sin\theta_{i} }
        { \sum_{i=1}^{3} (E_{i}^{+}\Delta\phi_{i}^{+}+E_{i}^{-}\Delta\phi_{i}^{-})\sin\theta_{i} }
\end{equation}
which are related to the amplitudes of the electric fields in Eqs.~\eqref{eq:trans-maxtrix}--\eqref{eq:refl}
and the phase factors in Eqs.~\eqref{eq:phase-fr}--\eqref{eq:phase-bk}.
It is noted that according to the conservation of the wave vector $k_{x}$ across each layer,
the nonlocal effective permittivity obtained from Eqs.~\eqref{eq:eps_x}--\eqref{eq:eps_z}
depends on both the frequency and the wave vector via $k_{x}=n_{i}k_{0}\sin\theta_{i}$
(where the wave vector of free space is $k_{0}=2\pi/\lambda$).
Therefore, both the frequency and the spatial dispersion are considered in the nonlocal EMT.

\section{Nonlocal EMT Analysis of Experimental Data}

In order to demonstrate the effects from the optical nonlocality, the symmetric $4$-pair $\mathrm{Ag}$-$\mathrm{SiO}_{2}$
multilayer stack on the top of a thick silver substrate shown in Fig.~\blue{1(a)} is then fabricated and the measured
reflection spectra are compared with the nonlocal EMT analysis.
The thickness of the $\mathrm{Ag}$ layer and of the $\mathrm{SiO}_{2}$ layer is designed to be $10\,\mathrm{nm}$
and $85\,\mathrm{nm}$, respectively.
The thickness of the silver substrate is $100\,\mathrm{nm}$ which is thick enough to block the optical transmission
in visible frequency region.
The multilayer stack is deposited on top of silicon substrates with the electron-beam evaporation system,
where $\mathrm{Ag}$ is deposited at the rate of $0.2$\,{\AA}$/\mathrm{sec}$
and $\mathrm{SiO}_{2}$ is deposited at $0.2$\,{\AA}$/\mathrm{sec}$.
Each material is individually deposited on a silicon substrate first to calibrate and optimize the deposition
parameters.
The optical constant of each material and the film thickness are characterized with the variable angle
spectroscopic ellipsometry (VASE, J. A. Woollam Co. VB400/HS-190).
The VASE measurements show that the optical constant of $\mathrm{Ag}$ matches the standard data of Johnson
and Christy \cite{Johnson1972PRB} based on the fitting from a general oscillator model.
The dielectric constant of $\mathrm{SiO}_{2}$ is fitted from the Sellmeier dispersion relation.
The VASE measured film thickness for each material also matches the thickness value for the set deposition parameters.
Figure~\blue{2} shows the scanning electron microscope (SEM) picture of the cross section of the
fabricated $\mathrm{Ag}$-$\mathrm{SiO}_{2}$ multilayer stack on the top of the silver substrate,
in which the focused ion beam (FIB) system (Helios Nanolab 600) is used to cut the cross section.
Each deposited thin layer can be clearly seen, where the bright and the dark stripes individually
correspond to the $\mathrm{Ag}$ layers and the $\mathrm{SiO}_{2}$ layers, together with the silver
substrate at the bottom.
The thickness of the deposited layers can be characterized with the VASE, and the measured averaged
thickness for the $\mathrm{Ag}$ layer, the $\mathrm{SiO}_2$ layer, and the silver substrate is
$10\pm0.4\,\mathrm{nm}$, $85\pm1.6\,\mathrm{nm}$, and $100\pm0.4\,\mathrm{nm}$, respectively.

The reflection spectra of the symmetric $\mathrm{Ag}$-$\mathrm{SiO}_2$ multilayer stack above the
silver substrate are then measured under the TM-polarized light in the wavelength range from $400\,\mathrm{nm}$
to $800\,\mathrm{nm}$, with respect to different angles of incidence from $0^{\circ}$ to $80^{\circ}$
with a variation of $10^{\circ}$.
The measured reflection spectra containing the information of optical nonlocality are compared
with the theoretical reflection spectra calculated from the multilayer stack, the local EMT,
and the nonlocal EMT at different angles of incidence, as shown in Fig.~\blue{3}.
It is clear that the theoretical reflection spectra obtained from the multilayer stack
calculation (black curves) are coincident with the experimental reflection spectra (red curves).
Meanwhile, the theoretical reflection spectra calculated from the nonlocal EMT (blue curves)
is also very close to the experimental data, showing the locations of the reflection minimums accurately.
However, without the consideration of optical nonlocality, the theoretical reflection spectra
obtained from the local EMT (dashed-blue curves) deviate far away from the experimental reflection
spectra with obvious blue shifts.

Furthermore, in order to indicate the effects of optical nonlocality on the effective permittivity,
Figs.~\blue{4(a)}--\blue{4(d)} plot the differences between the nonlocal effective permittivity and
the local effective permittivity defined as
$\Delta\varepsilon_{x} = \varepsilon_{x}^{\mathrm{nonloc}} - \varepsilon_{x}^{\mathrm{loc}}$
and
$\Delta\varepsilon_{z} = \varepsilon_{z}^{\mathrm{nonloc}} - \varepsilon_{z}^{\mathrm{loc}}$
for the $x$-component and the $z$-component of the effective permittivity, respectively.
It is shown that the $x$-component of the effective permittivity is more sensitive to the optical nonlocality,
since the variation of $\Delta\varepsilon_{x}$ is greater than the value of $\Delta\varepsilon_{z}$ in both
the real part and the imaginary part with respect to the variations of the wavelength and the angle of incidence.
In addition, the variation of the ENZ wavelength associated with the optical nonlocality (nonlocal ENZ wavelength)
that is determined as $\mathrm{Re}(\varepsilon_{x}^{\mathrm{nonloc}})=0$ is also plotted as the black curves
in Figs.~\blue{4(a)} and \blue{4(b)}, which indicate that the nonlocal ENZ wavelength varies from $628.693\,\mathrm{nm}$
to $612.268\,\mathrm{nm}$ as the angle of incidence changes from $0^{\circ}$ to $80^{\circ}$.
The difference between the nonlocal ENZ wavelength $\lambda_{\mathrm{ENZ}}^{\mathrm{nonloc}}$
and the local ENZ wavelength $\lambda_{\mathrm{ENZ}}^{\mathrm{loc}}=587.277\,\mathrm{nm}$ is also illustrated
in Fig.~\blue{4(e)} as a function of the angle of incidence.
The variation of the permittivity differences $\Delta\varepsilon_{x}$ and $\Delta\varepsilon_{z}$
at the local ENZ wavelength with respect to the angle of incidence are shown in Figs.~\blue{4(f)} and \blue{4(g)},
which reveal that the optical nonlocality has stronger influence on the $x$-component of the effective permittivity
than on the $z$-component of the effective permittivity.

The iso-frequency contours (IFCs) based on both the nonlocal EMT and the local EMT can be obtained
based on the nonlocal effective permittivity and the local effective permittivity,
\begin{equation}
\label{eq:nloc-ifc}
    \frac{k_{x}^{2}}{\varepsilon_{z}^{\mathrm{nonloc}}}
        + \frac{k_{z}^{2}}{\varepsilon_{x}^{\mathrm{nonloc}}} = k_{0}^{2}
\end{equation}
and
\begin{equation}
\label{eq:loc-ifc}
    \frac{k_{x}^{2}}{\varepsilon_{z}^{\mathrm{loc}}}
        + \frac{k_{z}^{2}}{\varepsilon_{x}^{\mathrm{loc}}} = k_{0}^{2}.
\end{equation}
The IFCs of the multilayer stack will also be calculated from the dispersion equation,
where the multilayer stack is considered as a one-dimensional photonic crystal \cite{Elser2007APL}
\begin{equation}
\label{eq:md-ifc}
    \cos(k_{z}(a_{m}+a_{d})) = \cos(k_{m}a_{m})\cos(k_{d}a_{d})
        -\frac{1}{2}\left( \frac{\varepsilon_{m}k_{d}}{\varepsilon_{d}k_{m}}
        + \frac{\varepsilon_{d}k_{m}}{\varepsilon_{m}k_{d}} \right)
        \sin(k_{m}a_{m})\sin(k_{d}a_{d}),
\end{equation}
in which $k_{m,d}^{2}=\varepsilon_{m,d}k_{0}^{2}-k_{x}^{2}$ for the TM-polarized light.
As displayed in Fig.~\blue{5}, the IFCs calculated from the nonlocal EMT in Eq.~\eqref{eq:nloc-ifc}
and the local EMT in Eq.~\eqref{eq:loc-ifc} are compared
with those obtained from the multilayer stack in Eq.~\eqref{eq:md-ifc} at three specific wavelengths:
the local ENZ wavelength
$\lambda_{\mathrm{ENZ}}^{\mathrm{loc}}=587.277\,\mathrm{nm}$ [Figs.~\blue{5(a)} and \blue{5(d)}],
the nonlocal ENZ wavelength associated with the $60^{\circ}$ angle of incidence
$\lambda_{\mathrm{ENZ},60^{\circ}}^{\mathrm{nonloc}}=616.255\,\mathrm{nm}$ [Figs.~\blue{5(b)} and \blue{5(e)}],
and the nonlocal ENZ wavelength related to the $0^{\circ}$ angle of incidence
$\lambda_{\mathrm{ENZ},0^{\circ}}^{\mathrm{nonloc}}=629.693\,\mathrm{nm}$ [Figs.~\blue{5(c)} and \blue{5(f)}].
The IFCs based on the nonlocal EMT [dashed curves in Figs.~\blue{5(a)}--\blue{5(c)}] are almost
the same as those of the multilayer stack (solid curves), especially in the area that the wave
vector $k_{x}$ is confined in the light cone of the air (green circles).
The deviation only occurs when the wave vector $k_{z}$ approaches to the boundary of the Brillouin zone,
where the multilayer stack cannot just be regarded as a homogenous and anisotropic effective medium since
the periodic property of the multilayer stack play important roles.
On the contrary, the IFCs based on the local EMT [dashed curves in Figs.~\blue{5(d)}--\blue{5(f)}]
are far away from the IFCs of the multilayer stack without considering the optical nonlocality.
Finally, it is notable that during the calculation, the values of the wave vectors are normalized
by the specified wave vector $k_{p}=4.56983\times10^{7}\,\mathrm{m}^{-1}$ that is related to the
$\mathrm{Ag}$ plasma frequency $\omega_{p}=1.37\times10^{16}\,\mathrm{rad}/\mathrm{s}$.

\section{Analysis of Dispersion Relation and Bulk Plasmon Mode}

The TM-polarized incident light will excite the SPPs at the metal-dielectric interfaces of the multilayer
stack when the wave vector $k_{x}$ extends beyond the light line of the free space.
Moreover, due to the coupling of the SPPs between different multilayer interfaces, the BPPs with large wave
vectors will be generated in the multilayer stack, which will strongly enhance the optical nonlocality.
The enhanced optical nonlocality can be illustrated by the differences between the nonlocal effective
permittivity and the local effective permittivity $\Delta\varepsilon_{x}$ and $\Delta\varepsilon_{z}$
with respect to the variations of the frequency and the wave vector, as shown in Fig.~\blue{6}.
Similar to the results in Figs.~\blue{4(a)}--\blue{4(d)}, $\Delta\varepsilon_{x}$ [Figs.~\blue{6(a)} and \blue{6(b)}]
is more sensitive to the optical nonlocality than $\Delta\varepsilon_{z}$ [Figs.~\blue{6(c)} and \blue{6(d)}]
in both the real part and the imaginary part.
Furthermore, the variation of the $\Delta\varepsilon_{x}$ is greatly increased when the wave vector $k_{x}$
extends to the region below the light line of free space, i.e., $k_{x}/k_{p}>\omega/\omega_{p}$ ,
where the BPPs are excited in the metal-dielectric multilayer stack and propagate along the
metal-dielectric interfaces in the $x$-direction.

In order to understand the mechanism for the giant $\Delta\varepsilon_{x}$ in the BPP region,
the dispersion relations calculated from the multilayer stack in Eq.~\eqref{eq:md-ifc},
the nonlocal EMT in Eq.~\eqref{eq:nloc-ifc} and the local EMT in Eq.~\eqref{eq:loc-ifc}
are investigated, as shown in Figs.~\blue{7(a)}, \blue{7(b)} and \blue{7(c)}, respectively.
The dispersion curves are separated by the light line of air (straight green line) into two regions.
Above the light line ($k_{x}/k_{p}<\omega/\omega_{p}$), the dispersion of the multilayer stack in
Fig.~\blue{7(a)} includes three branches, in which the first and the second branches
(\textit{A1} and \textit{A2}) located between
the experimental wavelength range from $400\,\mathrm{nm}$ to $800\,\mathrm{nm}$ are related to the
two minimums in the reflection spectra in Fig.~\blue{3}, while the third branch (\textit{A3})
is not observed in
the reflection spectra since it is outside the experimental wavelength range.
Below the light line ($k_{x}/k_{p}>\omega/\omega_{p}$), the dispersion of the multilayer stack
in Fig.~\blue{7(a)} possesses another four branches (\textit{B1}, \textit{B2}, \textit{B3}, and \textit{B4})
that are below the dispersion of the SPP (dashed-black curve) calculated via
$k_{x}=\sqrt{\varepsilon_{m}\varepsilon_{d}/(\varepsilon_{m}+\varepsilon_{d})}k_{0}$,
corresponding to four BPP modes associated with the four $\mathrm{Ag}$ layers in the multilayer stack.
Regarding to the dispersion relations calculated from the nonlocal EMT [Fig.~\blue{7(b)}]
and the local EMT [Fig.~\blue{7(c)}], it is shown that the three branches above the light
line are similar to those the multilayer stack in Fig.~\blue{7(a)}.
The three branches in the dispersion relation from the nonlocal EMT in Fig.~\blue{7(b)}
are almost the same as those from the multilayer stack, but the three branches from the
local EMT give slight blue shifts in Fig.~\blue{7(c)}.
However, the dispersion relations in the BPP region below the light line calculated from
the multilayer stack, the nonlocal EMT, and the local EMT are quite different.
According to the EMT, the multilayer stack is regarded as a bulk homogenous and anisotropic effective medium
so that there is no specific metal-dielectric interfaces inside the medium anymore during the procedure of
averaging the electromagnetic field across each periodic structure.
Therefore, the BPP modes supported in either the nonlocal medium in Fig.~\blue{7(b)}
or the local effective medium in Fig.~\blue{7(c)} are determined by the interference
of the SPPs between the two boundaries of the effective medium and therefore there are
many more branches shown in the BPP region.
With the consideration of the spatial dispersion related to the optical nonlocality
in the nonlocal EMT, the first four branches in the BPP region in the dispersion relation in Fig.~\blue{7(b)}
are almost identical with the four branches related to the BPP modes in the dispersion relation
of the multilayer stack in Fig.~\blue{7(a)}.
On the contrary, the dispersion relation from the local EMT does not perform such properties
in the BPP region, implying that the local EMT is not accurate to describe the BPP mode dispersion.
To further explore the dispersion relation of the multilayer stack, Fig.\blue{7(d)} illustrates
the eigenmodes of the electromagnetic field associated with the seven branches of the dispersion
relation in Fig.~\blue{7(a)}, in terms of the distributions of both the magnetic field amplitude $H_{y}$
and the magnetic field intensity $\left|H_{y}\right|$.
It is clear that the first three eigenmodes (\textit{A1}, \textit{A2}, and \textit{A3})
above the light line behaves as propagating modes along the $z$-direction in the multilayer
stack across through the whole multilayer stack.
On the other hand, the four BPP eigenmodes (\textit{B1}, \textit{B2}, \textit{B3}, and \textit{B4})
in the multilayer stack have the localized electromagnetic fields along the interfaces of the $\mathrm{Ag}$ layers
and the $\mathrm{SiO}_{2}$ layers due to the coupling of the SPP modes.

\section{Conclusions}

A nonlocal EMT has been derived based on the original definition of the effective permittivity
through the transfer-matrix method in order to study the optical nonlocality in the symmetric
metal-dielectric multilayer stack with respect to the TM-polarized incident light for different
incident angles.
Instead of the local EMT, the nonlocal EMT can accurately predict the measured incident
angle-dependent reflection spectra from the fabricated multilayer stack due to the
consideration of nonlocal effective permittivity.
The nonlocal EMT not only reveals the difference between the nonlocal effective permittivity
and the local effective permittivity but also the variation of IFCs induced by the optical nonlocality.
Furthermore, the BPP modes with large wave vectors enhanced by the optical nonlocality can also be
characterized with the nonlocal EMT.

\section*{Acknowledgments}

The authors acknowledge the support from the University of Missouri Interdisciplinary Intercampus
Research Program, the Ralph E. Powe Junior Faculty Enhancement Award, and the National Science
Foundation under grant CBET-1402743.
This work was performed, in part, at the Center for Integrated Nanotechnologies, an Office of
Science User Facility operated for the U.S. Department of Energy (DOE) Office of Science.
Sandia National Laboratories is a multi-program laboratory managed and operated by Sandia
Corporation, a wholly owned subsidiary of Lockheed Martin Corporation, for the U.S. Department
of Energy's National Nuclear Security Administration under contract DE-AC04-94AL85000.


\newpage                 %
\section*{Figure Captions}%
\noindent
\textbf{FIG.~1}. (Color online)
(a) Schematic of the symmetric and periodic $\mathrm{Ag}$-$\mathrm{SiO}_{2}$ multilayer stack on the top of the silver substrate.
(b) Schematic of the homogeneous and anisotropic effective medium on the silver substrate.
(c) Schematic of the symmetric unit cell of the $\mathrm{Ag}$-$\mathrm{SiO}_{2}$ multilayer stack with respect to
the TM-polarized incident light across each layer.

\vspace{5.0mm}
\noindent
\textbf{FIG.~2}. (Color online)
The SEM picture of the cross section of the fabricated $\mathrm{Ag}$-$\mathrm{SiO}_{2}$ multilayer stack on the silver substrate.

\vspace{5.0mm}
\noindent
\textbf{FIG.~3}. (Color online)
The measured reflection spectra from the $\mathrm{Ag}$-$\mathrm{SiO}_{2}$ multilayer stack (black curves),
the calculated reflection spectra based on the multilayer stack (red curves),
the nonlocal EMT (blue curves) and the local EMT (dashed-blue curves) at different
angles of incidence from $0^{\circ}$ to $80^{\circ}$ over the wavelength range from
$400\,\mathrm{nm}$ to $800\,\mathrm{nm}$.

\vspace{5.0mm}
\noindent
\textbf{FIG.~4}. (Color online)
The differences between the nonlocal effective permittivity and the local effective permittivity for
(a) $\mathrm{Re}(\Delta\varepsilon_{x})=\mathrm{Re}(\varepsilon_{x}^{\mathrm{nonloc}}-\varepsilon_{x}^{\mathrm{loc}})$,
(b) $\mathrm{Im}(\Delta\varepsilon_{x})=\mathrm{Im}(\varepsilon_{x}^{\mathrm{nonloc}}-\varepsilon_{x}^{\mathrm{loc}})$,
(c) $\mathrm{Re}(\Delta\varepsilon_{z})=\mathrm{Re}(\varepsilon_{z}^{\mathrm{nonloc}}-\varepsilon_{z}^{\mathrm{loc}})$,
and
(d) $\mathrm{Im}(\Delta\varepsilon_{z})=\mathrm{Im}(\varepsilon_{z}^{\mathrm{nonloc}}-\varepsilon_{z}^{\mathrm{loc}})$,
with respect to the variations of the angle of incidence and the wavelength.
(e) The differences between the nonlocal ENZ wavelength and the local ENZ wavelength as a function
of the angle of incidence.
The differences between the nonlocal effective permittivity and the local effective permittivity
with respect to the angle of incidence at the local ENZ wavelength for
(f) the $x$-component $\Delta\varepsilon_{x}$ and
(g) the $z$-component $\Delta\varepsilon_{z}$.

\vspace{5.0mm}
\noindent
\textbf{FIG.~5}. (Color online)
The comparison between the IFCs of the multilayer stack (solid curves) and the IFCs based on
the nonlocal EMT (dashed curves) at
(a) the local ENZ wavelength $\lambda_{\mathrm{ENZ}}^{\mathrm{loc}}=587.277\,\mathrm{nm}$,
(b) the nonlocal ENZ wavelength associated with the $60^{\circ}$ angle of incidence
$\lambda_{\mathrm{ENZ},60^{\circ}}^{\mathrm{nonloc}}=616.255\,\mathrm{nm}$, and
(c) the nonlocal ENZ wavelength related to the $0^{\circ}$ angle of incidence
$\lambda_{\mathrm{ENZ},0^{\circ}}^{\mathrm{nonloc}}=629.693\,\mathrm{nm}$.
The comparison between the IFCs of the multilayer stack (solid curves) and the IFCs based on
the local EMT (dashed curves) at
(c) the local ENZ wavelength $\lambda_{\mathrm{ENZ}}^{\mathrm{loc}}=587.277\,\mathrm{nm}$,
(d) the nonlocal ENZ wavelength associated with the $60^{\circ}$ angle of incidence
$\lambda_{\mathrm{ENZ},60^{\circ}}^{\mathrm{nonloc}}=616.255\,\mathrm{nm}$, and
(e) the nonlocal ENZ wavelength related to the $0^{\circ}$ angle of incidence
$\lambda_{\mathrm{ENZ},0^{\circ}}^{\mathrm{nonloc}}=629.693\,\mathrm{nm}$.

\vspace{5.0mm}
\noindent
\textbf{FIG.~6}. (Color online)
The differences between the nonlocal effective permittivity and the local effective permittivity for
(a) $\mathrm{Re}(\Delta\varepsilon_{x})=\mathrm{Re}(\varepsilon_{x}^{\mathrm{nonloc}}-\varepsilon_{x}^{\mathrm{loc}})$,
(b) $\mathrm{Im}(\Delta\varepsilon_{x})=\mathrm{Im}(\varepsilon_{x}^{\mathrm{nonloc}}-\varepsilon_{x}^{\mathrm{loc}})$,
(c) $\mathrm{Re}(\Delta\varepsilon_{z})=\mathrm{Re}(\varepsilon_{z}^{\mathrm{nonloc}}-\varepsilon_{z}^{\mathrm{loc}})$,
and
(d) $\mathrm{Im}(\Delta\varepsilon_{z})=\mathrm{Im}(\varepsilon_{z}^{\mathrm{nonloc}}-\varepsilon_{z}^{\mathrm{loc}})$
as a function of the wave vector $k_{x}/k_{p}$ and the frequency $\omega/\omega_{p}$.

\vspace{5.0mm}
\noindent
\textbf{FIG.~7}. (Color online)
The dispersion relation calculated from (a) the multilayer stack, (b) the nonlocal EMT, and (c) the local EMT.
(d) The eigenmodes of the electromagnetic field in the multilayer stack with respect to the dispersion relation (a)
in terms of the distributions of both the magnetic field amplitude $H_{y}$ and the magnetic field intensity $\left|H_{y}\right|$.


\clearpage
\newpage
\begin{figure}[htbp]
    \centering
    \includegraphics[width=98mm]{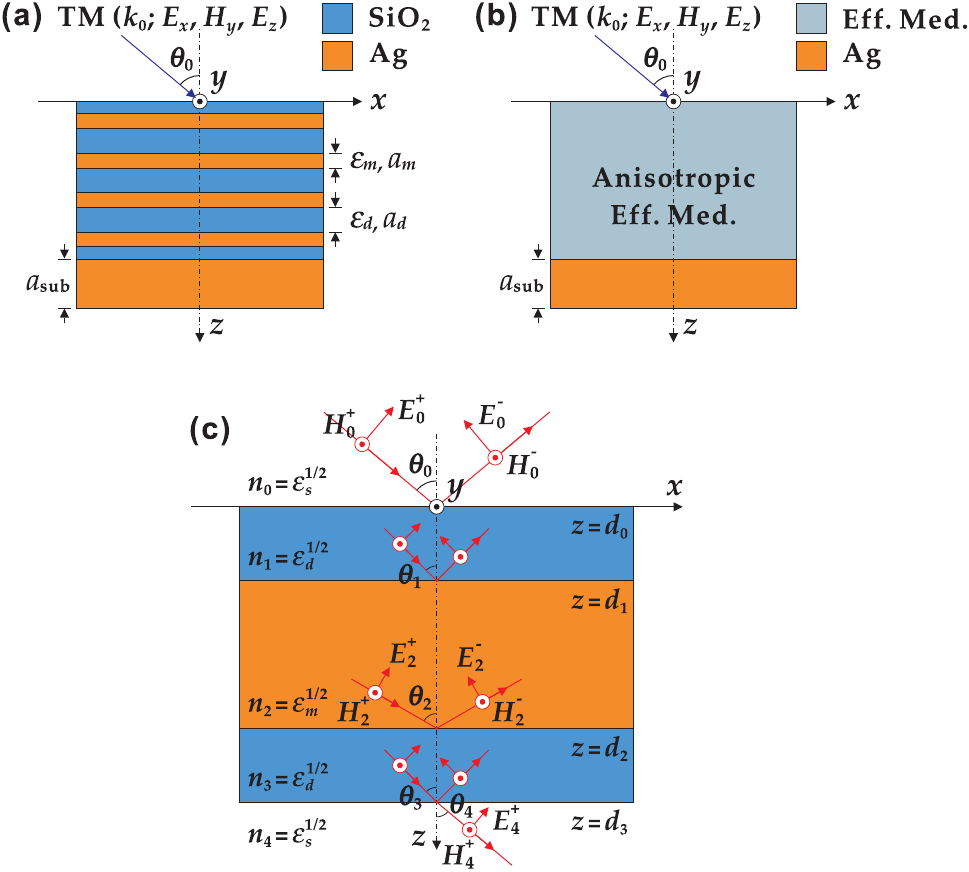}
    \caption{}
    \label{fig:fig1}
\end{figure}

\newpage
\begin{figure}[htbp]
    \centering
    \includegraphics[width=40mm]{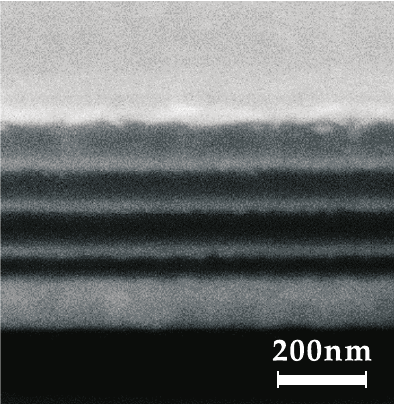}
    \caption{}
    \label{fig:fig2}
\end{figure}

\newpage
\begin{figure}[htbp]
    \centering
    \includegraphics[width=130mm]{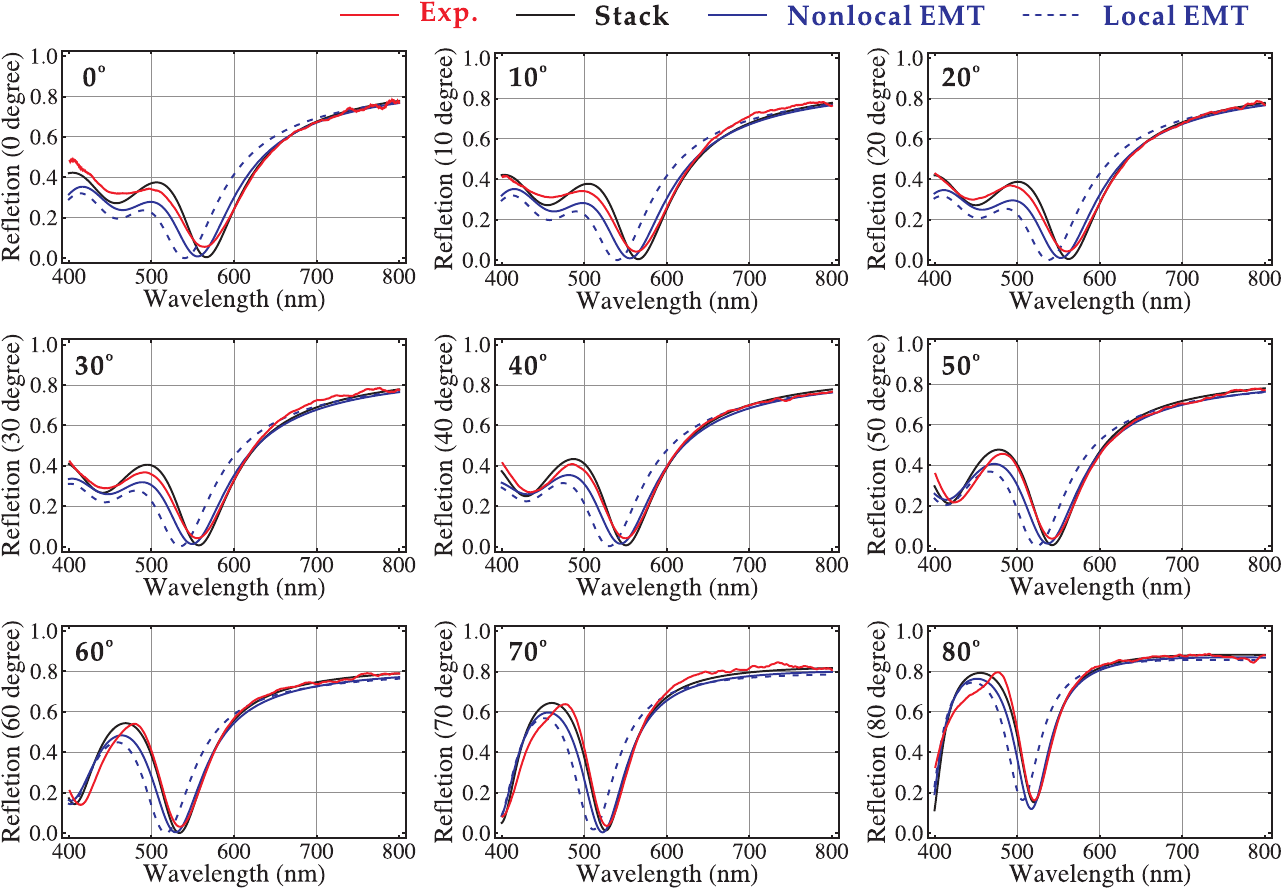}
    \caption{}
    \label{fig:fig3}
\end{figure}

\newpage
\begin{figure}[htbp]
    \centering
    \includegraphics[width=140mm]{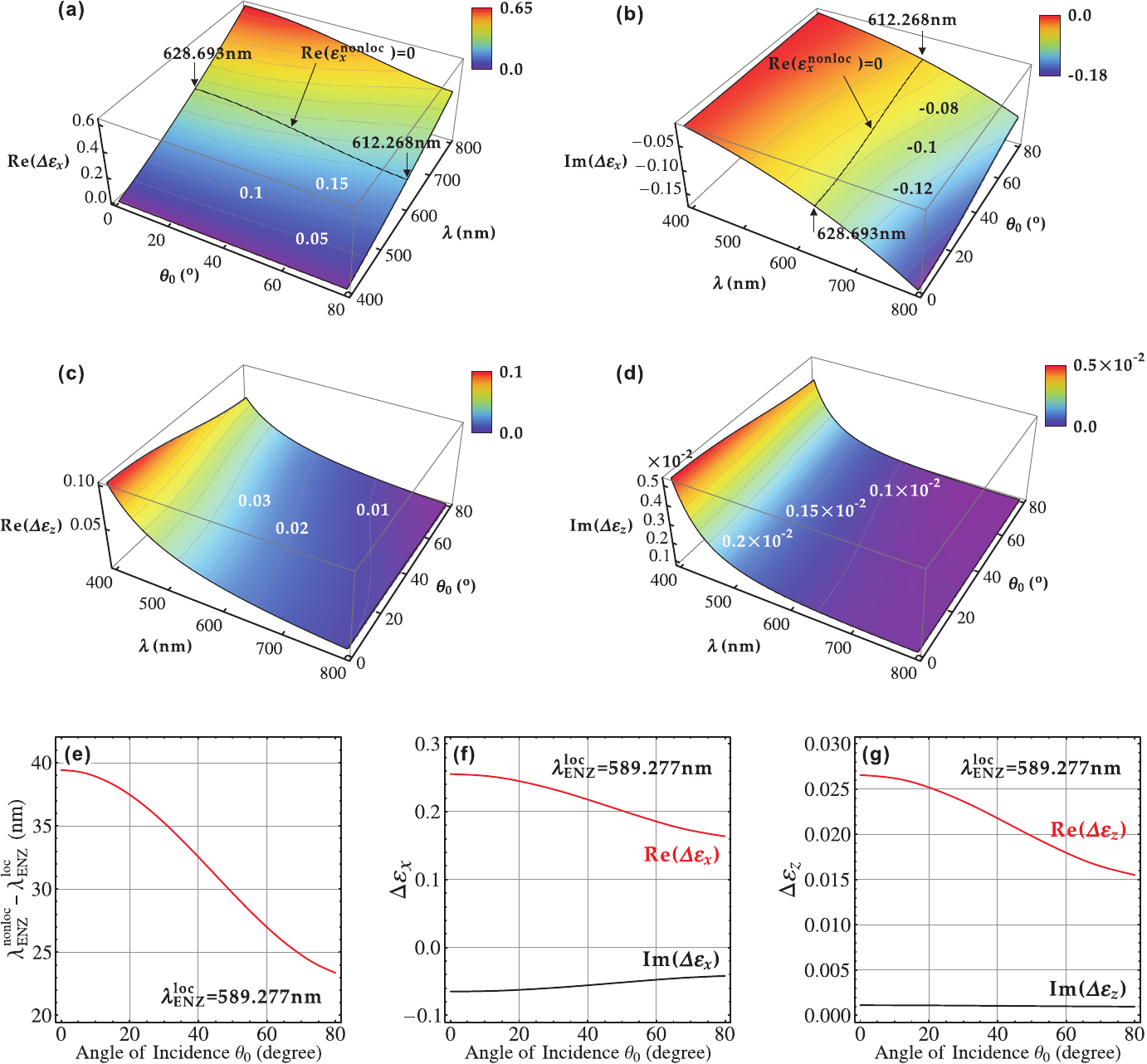}
    \caption{}
    \label{fig:fig4}
\end{figure}

\newpage
\begin{figure}[htbp]
    \centering
    \includegraphics[width=130mm]{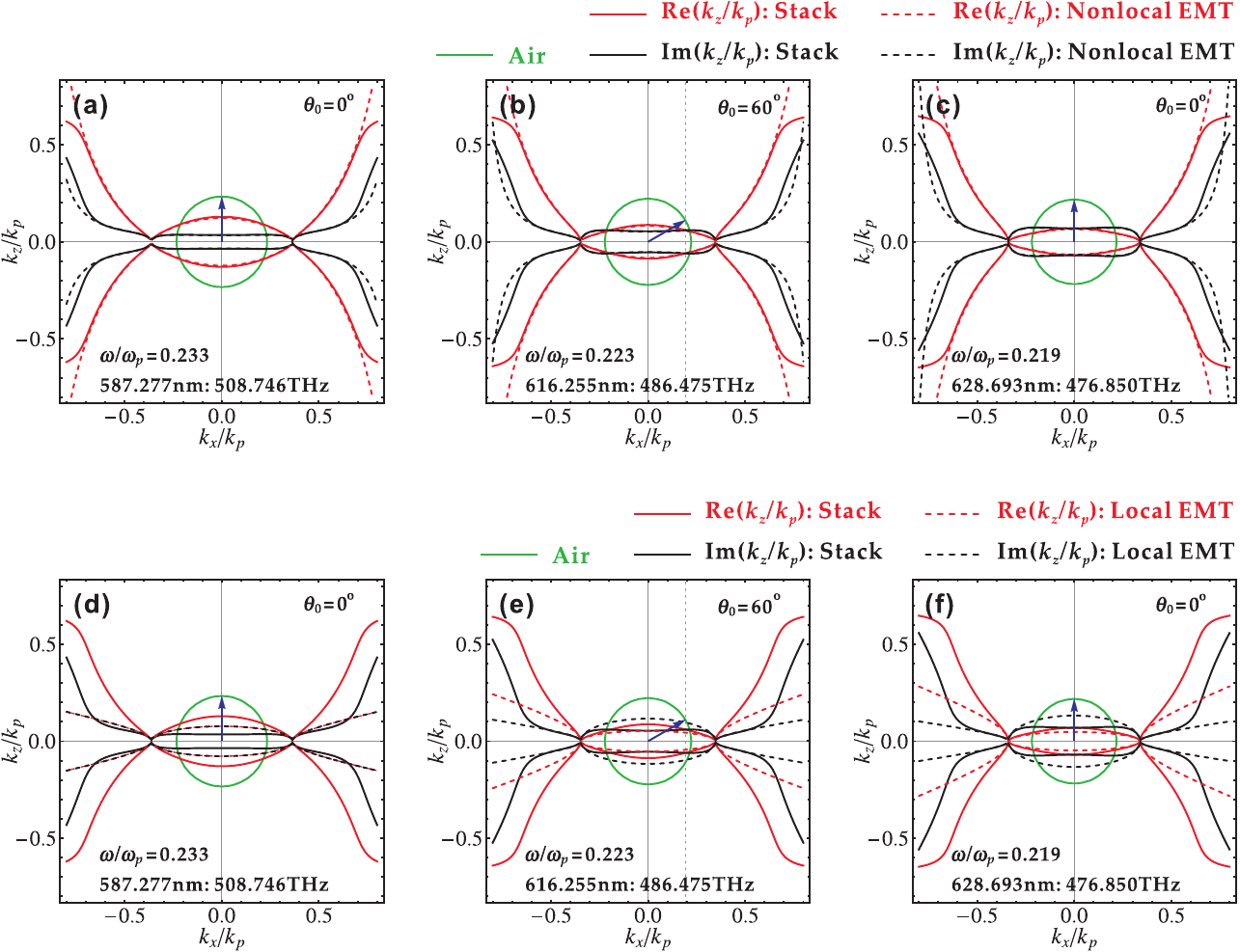}
    \caption{}
    \label{fig:fig5}
\end{figure}

\newpage
\begin{figure}[htbp]
    \centering
    \includegraphics[width=135mm]{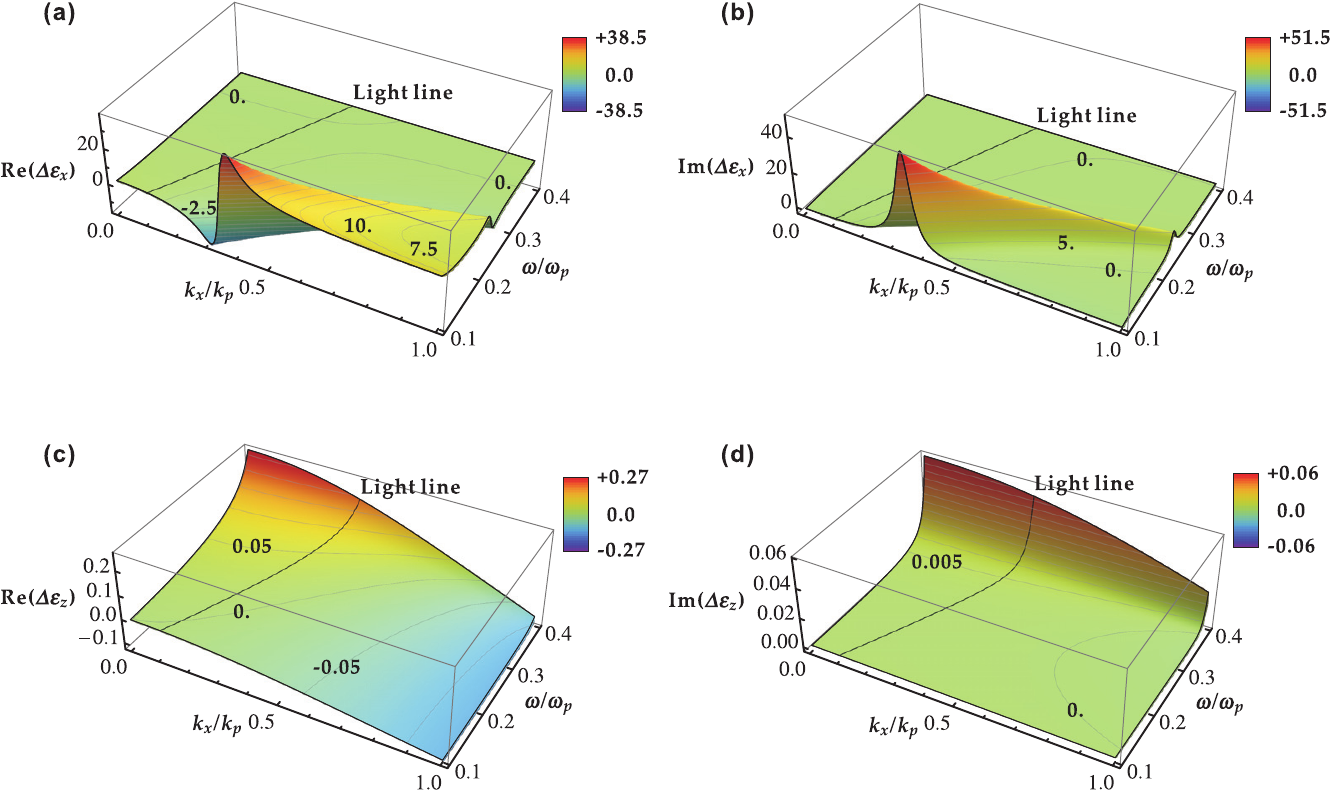}
    \caption{}
    \label{fig:fig6}
\end{figure}

\newpage
\begin{figure}[htbp]
    \centering
    \includegraphics[width=130mm]{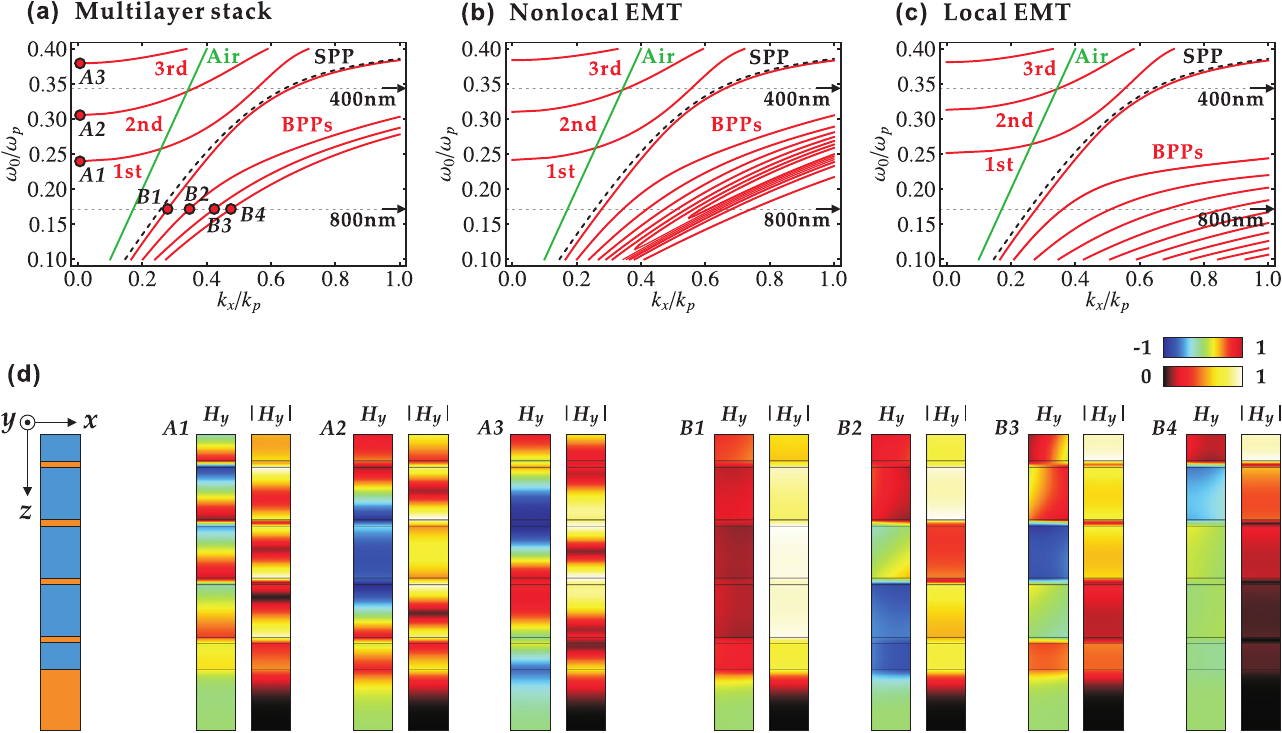}
    \caption{}
    \label{fig:fig7}
\end{figure}

\end{document}